\documentstyle[preprint,aps,epsf,floats]{revtex}

\begin{document}

\hbox to \hsize{\hfill
\vbox{\hbox{\bf MADPH-00-1182}
      \hbox{\bf FERMILAB-Pub-00/161-T}
      \hbox{\bf AMES-HET 00-07}
      \hbox{July 2000}}}

\vspace{.5in}

\begin{center}
{\large\bf Short-Baseline Neutrino Oscillations\\
at a Neutrino Factory}\\[6mm]
V. Barger$^1$, S. Geer$^2$, R. Raja$^2$, and K. Whisnant$^3$\\[3mm]
\it
$^1$Department of Physics, University of Wisconsin,
Madison, WI 53706, USA\\
$^2$Fermi National Accelerator Laboratory, P.O. Box 500,
Batavia, IL 60510, USA\\
$^3$Department of Physics and Astronomy, Iowa State University,
Ames, IA 50011, USA

\end{center}

\begin{abstract}

\vspace{-5ex}

Within the framework of three-neutrino and four-neutrino scenarios
that can describe the results of the LSND experiment, we consider the
capabilities of short baseline neutrino oscillation experiments at a
neutrino factory.  We find that, when short baseline ($L\alt 100$~km) neutrino
factory measurements are used together with other accelerator-based
oscillation results, the complete three-neutrino parameter space can
best be determined by measuring the rate of $\nu_e \to \nu_\tau$
oscillations, and measuring $CP$ violation with either $\nu_e \to
\nu_\mu$ or $\nu_\mu \to \nu_\tau$ oscillations (including the
corresponding antineutrino channels). With measurements of $CP$
violation in both $\nu_e \to \nu_\mu$ and $\nu_\mu \to \nu_\tau$ it may
be possible to distinguish between the three- and four-neutrino cases.

\end{abstract}

\thispagestyle{empty}

\vspace{5ex}

\section{Introduction}

The recent results from the Super-Kamiokande (SuperK) detector confirm the
solar~\cite{SuperKsolar} and atmospheric~\cite{SuperKatmos} neutrino deficits and strongly suggest the existence of neutrino oscillations. The azimuthal
angle and energy dependence of the atmospheric data  indicates
a mass-squared difference scale, $\delta m^2_{\rm atm} \sim
3.5\times10^{-3}$~eV$^2$.  The LSND measurements~\cite{LSND}
indicate neutrino oscillations with a different scale, $\delta m^2_{\rm
LSND} \sim 0.3$--$2.0$~eV$^2$; there is also a small region of acceptable parameter space at 6~eV$^2$.
The evidence for oscillations in solar neutrino
data~\cite{SuperKsolar,othersolar}, when taken as a whole, prefer yet a
third, lower mass-squared difference scale, $\delta m^2_{\rm sun} \le
10^{-4}$~eV$^2$. Once oscillation explanations for some or all of the
data are accepted, the next step is to attempt to find a neutrino mass
and mixing pattern that can provide a unified description of all the
relevant neutrino data.

In this paper we consider two possible neutrino scenarios which have
been proposed to account for the LSND results: (i) a three-neutrino
model that also describes the atmospheric neutrino data (in which case
the solar data would be explained by a phenomenon other than
oscillations~\cite{other}), and (ii) a four-neutrino model that also
describes both the solar and atmospheric data~\cite{bilenky,4nu}. As we
show, the three-neutrino model may be considered a sub-case of the
four-neutrino model. We examine the ability of short-baseline
experiments at a muon storage ring-based neutrino
factory~\cite{nufact,derujula,nufact2,BGW,BGRW,cervera,rubbia,freund,%
othernufact}
to determine the oscillation parameters in each case, and discuss how the three- and four-neutrino scenarios may be distinguished.

We concentrate on experiments in which muon and tau neutrinos are
detected via charge-current interactions and there is good sign
determination of the detected muons and tau leptons. The sign
determination allows one to distinguish between $\bar\nu_e \to
\bar\nu_\mu$  and $\nu_\mu \to \nu_\mu$ for stored $\mu^-$ in the
ring, and between $\nu_e \to \nu_\mu$ and $\bar\nu_\mu \to \bar\nu_\mu$
for stored $\mu^+$. We consider primarily the appearance channels
$\nu_e \to \nu_\mu$, $\nu_e \to \nu_\tau$, and $\nu_\mu \to \nu_\tau$
(and the corresponding antineutrino channels), for which the uncertainty on
the beam flux is not critical to the sensitivity of the
measurements. We find that the $\nu_\mu \to \nu_\tau$ and $\nu_e \to
\nu_\mu$ channels provide the best sensitivity to the $CP$-violating
phase; the $\nu_e \to \nu_\tau$ channel allows measurement of an
independent combination of mixing parameters. Short-baseline
experiments at a neutrino factory are sufficient to determine
the complete three-neutrino parameter space in these scenarios when their results are combined with other accelerator-based experiments
that are currently underway or being planned. Measuring $CP$-violation
in both $\nu_\mu \to \nu_\tau$ and $\nu_e \to \nu_\mu$ allows one to
possibly distinguish between the three- and four-neutrino cases. Hence
all three off-diagonal oscillation channels are useful. In the
four-neutrino case it is not possible to have complete mixing and
$\delta m^2$ parameter determinations without additional measurements
in future long-baseline experiments.

\section{Three-neutrino models}

Here we address the ability of short-baseline neutrino
oscillation experiments to probe a class of three-neutrino models that
can
describe the results of the LSND experiment, together with the
atmospheric neutrino deficit observed by the SuperK experiment.

\subsection{Oscillation Formalism}

In a three-neutrino model the neutrino flavor eigenstates $\nu_\alpha$
are related to the mass eigenstates $\nu_j$ in vacuum by a unitary
matrix $U$,
\begin{equation}
\left| \nu_\alpha \right> = \sum_j U_{\alpha j} \left| \nu_j
\right>\,,
\end{equation}
with $\alpha = e,\mu,\tau$ and $j=1,2,3$. The Maki-Nakagawa-Sakata
(MNS)\cite{mns} mixing matrix can be parameterized by
\begin{equation}
U
= \left( \begin{array}{ccc}
  c_{13} c_{12}       & c_{13} s_{12}  & s_{13} e^{-i\delta} \\
- c_{23} s_{12} - s_{13} s_{23} c_{12} e^{i\delta}
& c_{23} c_{12} - s_{13} s_{23} s_{12} e^{i\delta}
& c_{13} s_{23} \\
    s_{23} s_{12} - s_{13} c_{23} c_{12} e^{i\delta}
& - s_{23} c_{12} - s_{13} c_{23} s_{12} e^{i\delta}
& c_{13} c_{23} \\
\end{array} \right) \,,
\label{eq:MNS}
\end{equation}
where $c_{jk} \equiv \cos\theta_{jk}$, $s_{jk} \equiv \sin\theta_{jk}$,
and $\delta$ is the $CP$ non-conserving phase. Two additional diagonal
phases are present in $U$ for Majorana neutrinos, but these do not
affect oscillation probabilities.

\begin{figure}
\centering\leavevmode
\epsfxsize=4in\epsffile{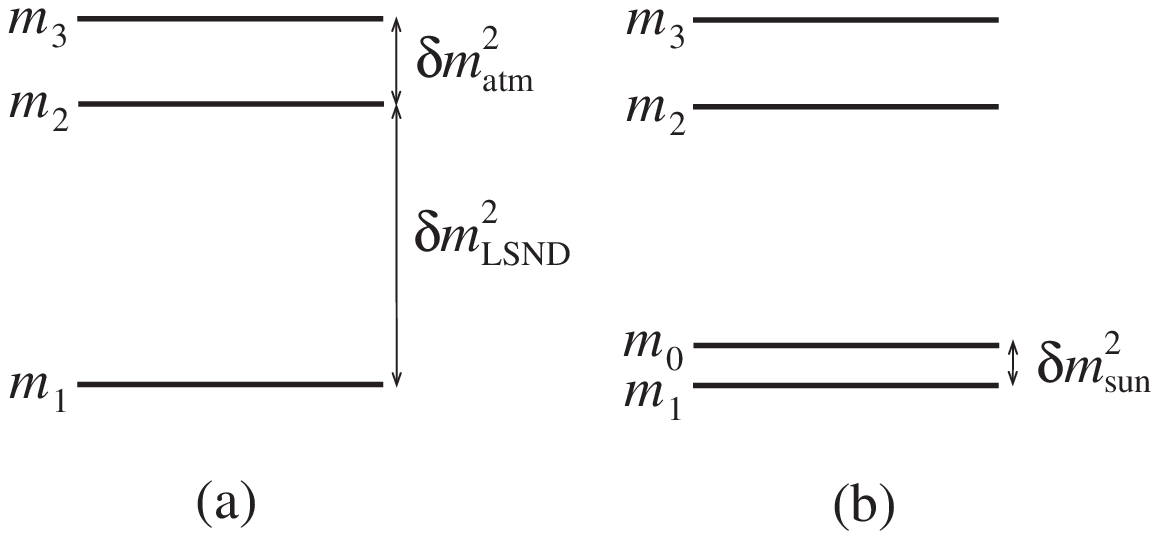}

\bigskip

\caption[]{Ordering and separation of mass eigenvalues in the (a)
three-neutrino and (b)~four-neutrino scenarios in this paper.}
\label{fig:levels}
\end{figure}

For three-neutrino models in which the results of the
atmospheric and LSND experiments are explained by neutrino oscillations
there are two independent mass-squared differences:
$\delta m^2_ {\rm LSND} = 0.3$--$2.0$~eV$^2$ and
$\delta m^2_ {\rm atm} \simeq 3.5\times10^{-3}$~eV$^2$.
We take $\delta m^2_{32} = \delta m^2_{\rm atm}$ and
$\delta m^2_{21} = \delta m^2_{\rm LSND}$; see Fig.~\ref{fig:levels}a.
These mass-squared differences obey the condition
$\delta m^2_{\rm LSND} \gg \delta m^2_{\rm atm}$. Then
the general off-diagonal vacuum oscillation probability is
\begin{equation}
P(\nu_\alpha \to \nu_\beta) \simeq
4|U_{\alpha1}|^2|U_{\beta1}|^2 \sin^2\Delta_{\rm LSND}
-4 {\rm Re} (U_{\alpha2} U_{\alpha3}^* U_{\beta2}^* U_{\beta3})
\sin^2\Delta_{\rm atm}
\pm 2JS \,,
\label{eq:prob}
\end{equation}
where
\begin{eqnarray}
\Delta_{j} &\equiv&
1.27 \delta m^2_{j} ({\rm eV}^2) L ({\rm km}) / E_\nu ({\rm GeV}) \,,
\\
S &\equiv& \left[ \sin2\Delta_{\rm atm} + \sin2\Delta_{\rm LSND}
- \sin2(\Delta_{\rm LSND}+\Delta_{\rm atm}) \right]
\nonumber\\
&=& 2 (\sin2\Delta_{\rm atm}\sin^2\Delta_{\rm LSND}
+ \sin2\Delta_{\rm LSND}\sin^2\Delta_{\rm atm}) \,,
\label{eq:S}
\end{eqnarray}
and $J$ is the $CP$-violating invariant\cite{keung,jarlskog}, which
can be defined as $J = Im\{ U_{e2} U_{e3}^* U_{\mu 2}^* U_{\mu 3}\}$.
The plus (minus) sign in Eq.~(\ref{eq:prob}) is used when $\alpha$ and
$\beta$ are in cyclic (anticyclic) order, where cyclic order is defined
as $e\mu\tau$. For antineutrinos, the sign of the $CP$-violating term
is reversed. For the mixing matrix in Eq.~(\ref{eq:MNS}),
\begin{equation}
J = s_{13} c_{13}^2 s_{12} c_{12} s_{23} c_{23} \sin\delta \,.
\label{eq:J}
\end{equation}
For recent discussions of $CP$ violation in neutrino oscillations, see
Refs.~\cite{derujula}, \cite{BGRW}--\cite{rubbia} and
\cite{4nuCP}--\cite{CPV}. For $L \le 100$~km, matter
effects~\cite{matter} are very small and the vacuum formulas are a good
approximation to the true oscillation probabilities. 

The class of scenarios that we are considering is designed to account
for the large $\nu_\mu \to \nu_\tau$ mixing of atmospheric neutrinos and
the small $\nu_\mu \to \nu_e$ mixing in the LSND experiment. The $\nu_e$
survival probability at the leading oscillation scale is given by
\begin{equation}
P(\nu_e \to \nu_e) = P(\bar\nu_e \to \bar\nu_e)
= 4 c_{12}^2 c_{13}^2 (1 - c_{12}^2 c_{13}^2) \sin^2\Delta_{\rm LSND} \,.
\end{equation}
Results from the BUGEY reactor experiment~\cite{BUGEY} put an upper
bound on $P(\bar\nu_e \to \bar\nu_e)$ for $\delta m^2_{\rm LSND} >
0.01$~eV$^2$,  which provides the approximate constraint 
\begin{equation}
s_{12}^2 + s_{13}^2 \le 0.01 \,.
\end{equation}
This leads to the conditions
\begin{equation}
\theta_{12}, \theta_{13} \ll \theta_{23} \,,
\label{eq:small}
\end{equation}
with the mixing of atmospheric neutrinos, $\theta_{23}$, near maximal
($\theta_{23} \sim \pi/4$). Hence we can take $s_{12}, s_{13} \ll s_{23},
c_{23}$ and $c_{12} \simeq c_{13} \simeq 1$. The
off-diagonal oscillation probabilities are (to leading order in the
small mixing angle parameters)
\begin{eqnarray}
P(\nu_e \to \nu_\mu) &\simeq&
4|s_{12}c_{23}+s_{13}s_{23}e^{i\delta}|^2 \sin^2\Delta_{\rm LSND}
- 2 s_{12}s_{13}c_\delta\sin2\theta_{23} \sin^2\Delta_{\rm atm}
\nonumber\\
&\phantom{\simeq}&
+ s_{12}s_{13}s_\delta\sin2\theta_{23} S \,,
\label{eq:Pem}\\
P(\nu_e \to \nu_\tau) &\simeq&
4|s_{12}s_{23}-s_{13}c_{23}e^{i\delta}|^2 \sin^2\Delta_{\rm LSND}
+ 2 s_{12}s_{13}c_\delta\sin2\theta_{23} \sin^2\Delta_{\rm atm}
\nonumber\\
&\phantom{\simeq}&
 - s_{12}s_{13}s_\delta\sin2\theta_{23} S \,,
\label{eq:Pet}\\
P(\nu_\mu \to \nu_\tau) &\simeq&
4|s_{12}c_{23}+s_{13}s_{23}e^{i\delta}|^2
|s_{12}s_{23}-s_{13}c_{23}e^{-i\delta}|^2 \sin^2\Delta_{\rm LSND}
\nonumber\\
&\phantom{\simeq}&
+ \sin^22\theta_{23} \sin^2\Delta_{\rm atm}
+ s_{12}s_{13}s_\delta\sin2\theta_{23} S \,,
\label{eq:Pmt}
\end{eqnarray}
where $c_\delta = \cos\delta$ and $s_\delta = \sin\delta$. The
expressions for the antineutrino channels are obtained by changing the
sign of the $S$-term in each case. A representative scenario, 1B1 in
Ref.~\cite{1B1}, that we will use as an example has the parameters
\begin{eqnarray}
\delta m^2_{32} &=& 3.5\times10^{-3}{\rm~eV}^2 \,,
\sin^22\theta_{23} = 1.0
\nonumber\\
\delta m^2_{21} &=& 0.3{\rm~eV}^2 \,,
\sin^22\theta_{12} = \sin^22\theta_{13} = 0.015 \,,
\label{eq:1B1}
\end{eqnarray}
with $\delta$ a varied parameter. More generally, $\delta m^2_{\rm
LSND}$ in the range 0.3--2.0~eV$^2$ is allowed with $\nu_\mu \to \nu_e$
oscillation amplitude given by
\begin{equation}
4|s_{12}c_{23}+s_{13}s_{23}e^{i\delta}|^2 \simeq
\left({0.06{\rm~eV}^2\over \delta m^2_{\rm LSND}} \right)^2 \,.
\label{eq:LSNDamp}
\end{equation}

In a short-baseline experiment, $L/E$ should optimally be chosen such
that
$\Delta_{\rm LSND} \sim 1$, in which case $\Delta_{\rm atm} \simeq
\Delta_{\rm LSND}/100 \ll 1$; e.g., for $\delta m^2_{\rm LSND} =
0.3$~eV$^2$ and $E_\nu = 14$~GeV we might choose $L = 45$~km, giving:
\begin{equation}
\Delta_{\rm atm} = 0.014
\left({\delta m^2_{\rm atm}\over3.5\times10^{-3}{\rm~eV}^2}\right)
\left({L\over45{\rm~km}}\right)
\left({14 {\rm~GeV}\over E_\nu}\right) \,.
\end{equation}
Note that in the forward direction $E_\nu = 14$~GeV is the average
$\nu_\mu$ energy for stored unpolarized $\mu^-$ with $E_\mu = 20$~GeV.
Thus in the probability equations above,
$s_{12}$, $s_{13}$, and $\Delta_{\rm atm}$ are all small parameters, at
the few percent level or less. Then to leading order in $\Delta_{\rm
atm}$
\begin{equation}
S \simeq 4 \Delta_{\rm atm} \sin^2\Delta_{\rm LSND} \,.
\label{eq:Sapprox}
\end{equation}
Therefore, in these
scenarios, the dominant contribution to $P(\nu_e \to \nu_\mu)$ and
$P(\nu_e \to \nu_\tau)$ comes from the leading oscillation
(the $\sin^2\Delta_{\rm LSND}$ term), but the dominant contribution to
$P(\nu_\mu \to \nu_\tau)$ comes from the subleading oscillation (the
$\sin^2\Delta_{\rm atm}$ term); these results are summarized in
Table~\ref{tab:CP}. In each case, the dominant term is $CP$-conserving
and proportional to the product of two small parameters. The
$CP$-violating contribution in each case is $2s_{12}s_{13}s_\delta
\sin2\theta_{23}\sin2\Delta_{\rm atm}\sin^2\Delta_{\rm LSND}$,
which is a product of three small parameters (assuming $\delta$ is not
small), and hence is smaller than the $CP$-conserving
contribution (see Table~\ref{tab:CP}).

\begin{table}[h]
\caption{Leading and next-to-leading contributions to neutrino
oscillations in the three-neutrino models under consideration in a
short-baseline experiment with $L/E$ chosen such that
$\Delta_{\rm LSND} \sim 1$. The parameters $s_{12}$, $s_{13}$ and
$\Delta_{\rm atm} \simeq \Delta_{\rm LSND}/100$ are all $\ll 1$. Also
shown is the size of the $CP$-violating ($CPV$) term compared
to the (dominant) $CP$-conserving ($CPC$) term, assuming terms the same
order of magnitude cancel in the ratio. \label{tab:CP}}
\def\arraystretch{1.1}
\begin{tabular}{cccc}
Oscillation & Leading term & Next-to-leading term & Ratio of $CPV$\\
& ($CP$ conserving) & ($CP$-violating) & to $CPC$ terms\\
\hline
$P(\nu_e \to \nu_\mu$)
& $4|s_{12}c_{23}+s_{13}s_{23}e^{i\delta}|^2 \sin^2\Delta_{\rm LSND}$
& $4s_{12}s_{13}s_\delta\sin2\theta_{23}\sin^2\Delta_{\rm LSND}\Delta_{\rm
atm}$
& $s_\delta\Delta_{\rm atm}$\\
$P(\nu_e \to \nu_\tau$)
& $4|s_{12}s_{23}-s_{13}c_{23}e^{i\delta}|^2 \sin^2\Delta_{\rm LSND}$
& $-4s_{12}s_{13}s_{\delta}\sin2\theta_{23}\sin^2\Delta_{\rm LSND}\Delta_{\rm
atm}$
& -$s_\delta\Delta_{\rm atm}$\\
$P(\nu_\mu \to \nu_\tau)$ & $\sin^22\theta_{23}\Delta_{\rm atm}^2$
& $4s_{12}s_{13}s_{\delta}\sin2\theta_{23}\sin^2\Delta_{\rm LSND}\Delta_{\rm atm}$
& ${s_{12}s_{13}s_\delta\over\Delta_{\rm atm}}$
\end{tabular}
\end{table}

In all, there are six parameters to be determined in the expressions in
Eqs.~(\ref{eq:Pem})--(\ref{eq:Pmt}): the three mixing angles, the phase
$\delta$, 
and two independent mass-squared differences. However, although there
are six off-diagonal measurements possible with $\mu$ and $\tau$
detection [the three in Eqs.~(\ref{eq:Pem})--(\ref{eq:Pmt}) plus the
corresponding antineutrino channels], the leading and next-to-leading
terms in the expressions for these oscillation probabilities are only
sensitive to five independent quantities:
$|s_{12}c_{23}+s_{13}s_{23}e^{i\delta}|$,
$|s_{12}s_{23}-s_{13}c_{23}e^{i\delta}|$, $s_{12}s_{13}s_{\delta}$,
$\sin^2\Delta_{\rm LSND}$, and $\sin2\theta_{23}\Delta_{\rm atm}$.
For the parameter ranges we are considering, $\theta_{12}, \theta_{13}
> \Delta_{\rm atm}$; then the $\sin^2\Delta_{\rm LSND}$ term in
Eq.~(\ref{eq:Pmt}) can be comparable to one of the other terms in
$P(\nu_\mu \to \nu_\tau)$, but this term still depends
on a subset of these same five independent quantities.


The K2K~\cite{K2K}, MINOS~\cite{MINOS}, ICANOE~\cite{ICANOE}
and OPERA~\cite{OPERA} long-baseline experiments will measure the
parameters in the leading term of the $\nu_\mu \to \nu_\tau$
probability, $\theta_{23}$ and $\delta m^2_{\rm atm}$, and
MiniBooNE~\cite{MiniBooNE} will measure the parameters in the leading
term of the $\nu_e \to \nu_\mu$ probability, $|s_{12}c_{23} +
s_{13}s_{23}e^{i\delta}|$ and $\delta m^2_{\rm LSND}$. Therefore, only
two independent quantities will remain to be measured in short baseline
experiments: (i) the amplitude of the leading oscillation in the $\nu_e
\to \nu_\tau$ probability, and (ii) the subleading CPV term, which has
the same magnitude for each off-diagonal channel and can be determined
by a comparison of neutrino to antineutrino rates. Hence a combination
of short-baseline measurements with the results of the other
accelerator-based experiments would allow all of the parameters in this
three-neutrino scenario to be determined. Measurements of the other
short-baseline off-diagonal oscillation probabilities may then be used
to check the consistency of the result and/or improve the accuracy of
the parameter determinations.

One can also measure the $\nu_\mu \to \nu_\mu$ survival probability,
which to leading order in small quantities can be written
\begin{equation}
P(\nu_\mu \to \nu_\mu) \simeq 1 - 4 A (1-A) \sin^2\Delta_{\rm LSND}
-\sin^22\theta_{23}\sin^2\Delta_{\rm atm} \,,
\label{eq:Pmm}
\end{equation}
where
\begin{equation}
A \simeq s_{23}^2s_{13}^2 + c_{23}^2s_{12}^2
+ 2 s_{23}c_{23}s_{12}s_{13}c_\delta \,.
\end{equation}
In short-baseline experiments, both oscillatory terms in
Eq.~(\ref{eq:Pmm}) are second order in small quantities. A measurement
of $P(\nu_\mu \to \nu_\mu)$ could be used in conjunction with
other short-baseline measurements to make a complete determination of
the oscillation parameters without the use of other data. However,
since $A$ and $\Delta_{\rm atm}$ are small, the deviations of
$P(\nu_\mu \to \nu_\mu)$ from unity are also small, and the
normalization of the beam flux would need to be known to high precision
for this to be a useful measurement.

An examination of Eqs.~(\ref{eq:Pem})--(\ref{eq:Pmt}) and
(\ref{eq:Sapprox}) shows that only the relative sign of $\delta$ and
$\delta m^2_{32}$ may be determined in short-baseline measurements for
the distances and oscillation parameter values that we are
considering. Long-baseline experiments at a neutrino
factory should be able to determine the sign of $\delta
m^2_{32}$ since there is a significant dependence of the matter effect
on the sign of $\delta m^2_{32}$ for $L \ge
2000$~km~\cite{BGRW,cervera,freund}.

\subsection{Results}

\begin{figure}
\centering\leavevmode
\epsfxsize=4.5in\epsffile{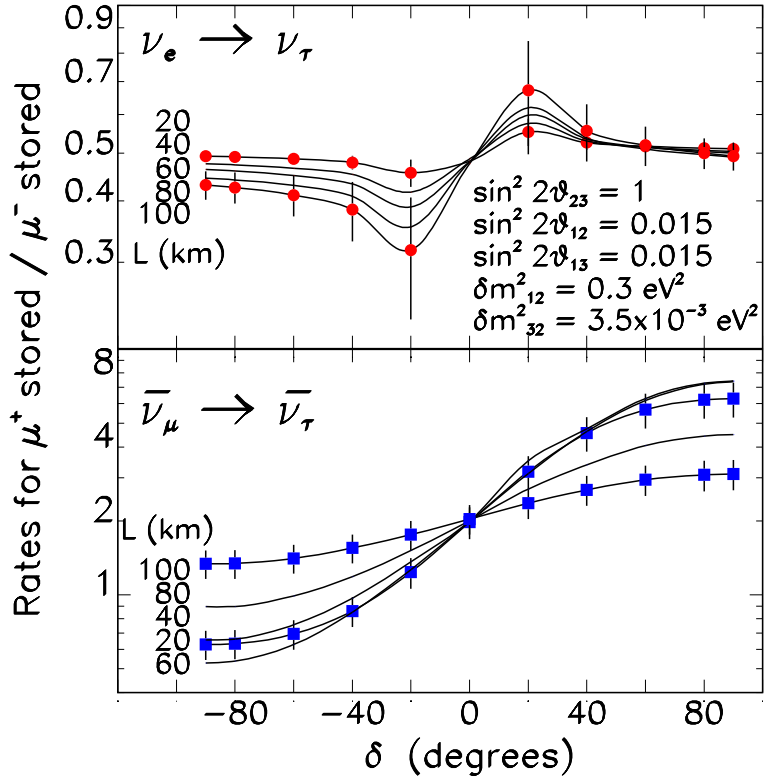}

\bigskip

\caption[]{Ratio of tau production rates for stored $\mu^+$ to that with
stored $\mu^-$ for $\nu_e \to \nu_\tau$ and $\nu_\mu \to \nu_\tau$
oscillations, shown versus the $CP$-violating phase $\delta$ for
several values of $L$. These results assume a stored muon energy $E_\mu
= 20$~GeV and oscillation parameters given by Eq.~(\ref{eq:1B1}). The
representative errors shown are statistical, assuming
$10^{20}$~kt-decays (after accounting for detector efficiency).}
\label{fig:combo}
\end{figure}

Figure~\ref{fig:combo}
shows the ratio of the rate of tau production for stored $\mu^+$ to that
for stored $\mu^-$ for the $\nu_e \to \nu_\tau$ and $\nu_\mu \to
\nu_\tau$ channels, versus the $CP$ phase $\delta$ for several values of
baseline length $L$, with oscillation parameters given by
Eq.~(\ref{eq:1B1}). Statistical errors are also shown, assuming
$10^{20}$ kt-decays (corresponding, for example, to
three years of running with $10^{20}$ useful muon decays per year
and a 1~kt detector having 33\% tau detection
efficiency). The event rate calculations are done according to the
method outlined in Ref.~\cite{BGW}.
For the choice of parameters in Eq.~(\ref{eq:1B1}), $s_{12}$ and
$s_{13}$ are larger than $\Delta_{\rm atm}$ and the $\nu_\mu \to
\nu_\tau$ channel shows the largest relative $CP$-violating effect
(see the last column of Table~\ref{tab:CP}).

The last column in Table~\ref{tab:CP} shows that the relative size of
the $CP$-violating effect in the $\nu_\mu \to \nu_\tau$ channel
decreases with increasing $\Delta_{\rm atm}$ (i.e., with increasing
$L/E$). Also, for very small $L/E$, when $\Delta_{\rm LSND} \ll 1$, $S
\to 0$ to leading order in $\Delta_{\rm atm}$ and $\Delta_{\rm LSND}$,
and the $CP$ violation becomes negligible.  Therefore, for any given
set of oscillation parameters, there will be an optimum $L/E$ that
maximizes the $CP$ violation effects in the $\nu_\mu \to \nu_\tau$
channel.

\begin{figure}
\centering\leavevmode
\epsfxsize=6.3in\epsffile{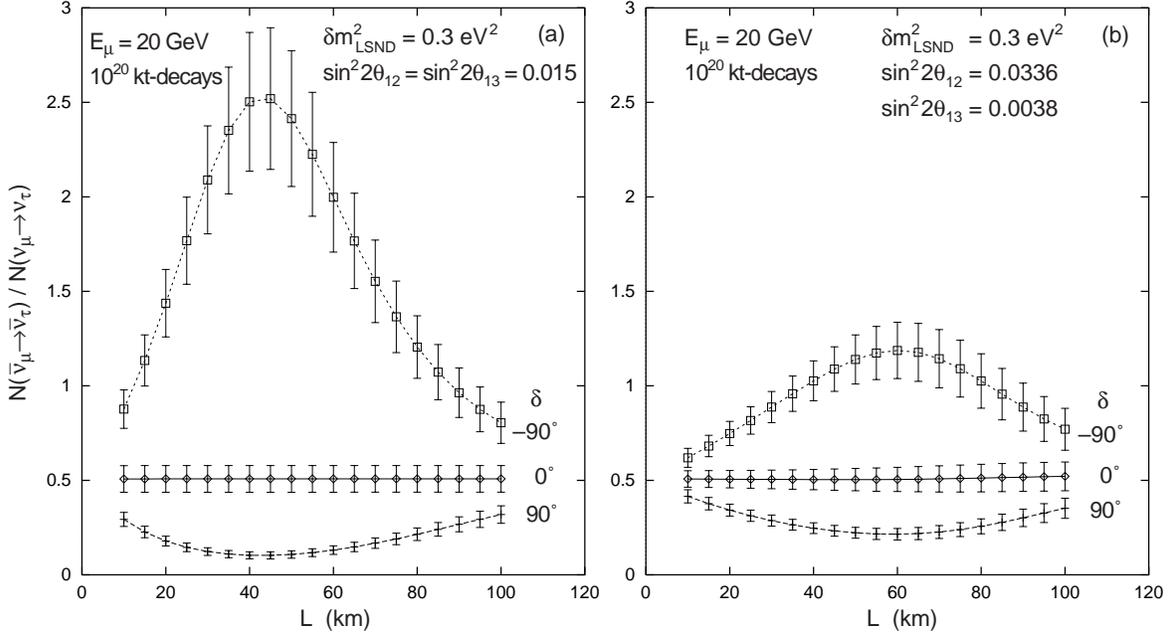}

\bigskip

\caption[]{(a) $R_{\mu\tau} \equiv N(\bar\nu_\mu \to
\bar\nu_\tau)/N(\nu_\mu \to \nu_\tau)$ versus $L$, for $E_\mu = 20$~GeV,
with oscillation parameters given by
Eq.~(\ref{eq:1B1}), where $\theta_{12} = \theta_{13}$. (b) Similar
results for $\sin^22\theta_{12} = 0.0336$ and $\sin^22\theta_{13} =
0.0038$. Statistical errors correspond to $10^{20}$~kt-decays.}
\label{fig:mutau}
\end{figure}

Figure~\ref{fig:mutau}a shows the ratio of $\bar\nu_\mu \to
\bar\nu_\tau$ event rates (from $\mu^+$ decays) to $\nu_\mu \to \nu_\tau$
event rates (from $\mu^-$ decays), $R_{\mu\tau}$, versus baseline for
20~GeV muons, for oscillation parameters given by Eq.~(\ref{eq:1B1}) and
three values of $\delta$ ($90^\circ$, $0^\circ$, and $-90^\circ$).
Approximate statistical errors are shown for
$10^{20}$~kt-decays. The figure clearly shows that there is one
distance that maximizes the size of the $CP$-violation effect, which in
this case (20~GeV muons) is about $L=45$~km. This optimal distance
decreases slowly with increasing $\delta m^2_{32}$. For $\delta
m^2_{32}$ in the range $2.5$--$4.5\times10^{-3}$~eV$^2$, we find that
the optimal $L$ is in the range $40$--$50$~km for $\delta m^2_{\rm LSND}
= 0.3$~eV$^2$. The optimal $L$ scales inversely with $\delta m^2_{\rm
LSND}$, and for stored muon energies well above the tau threshold may be
approximated by
\begin{equation}
L_{\rm opt} \simeq 45{\rm~km}
\left( {0.3{\rm~eV}^2\over\delta m^2_{\rm LSND}} \right)
\left( {E_\mu\over20{\rm~GeV}} \right) \,;
\label{eq:Lopt}
\end{equation}
e.g., for $\delta m^2_{\rm LSND} = 2$~eV$^2$ the best sensitivity to
$CP$ violation is obtained with $L \simeq 6$~km. The distance from
Fermilab to Argonne is about 30~km, which would be optimal for
$E_\mu=20$~GeV and $\delta m^2_{\rm LSND} =0.45$~eV$^2$. Similar
results for an optimal distance for $CP$ violation in $\nu_\mu \to
\nu_\tau$ in the context of four-neutrino models have been reported
in Ref.~\cite{donini}.

Given Eq.~(\ref{eq:LSNDamp}) (the LSND constraint on the $\nu_\mu \to
\nu_e$ oscillation amplitude), $CP$ violation is maximized when
$\theta_{12} = \theta_{13}$ since $J$ is proportional to the product of
$s_{12}$ and $s_{13}$. Figure~\ref{fig:mutau}b shows $R_{\mu\tau}$
for unequal $\theta_{12}$ and $\theta_{13}$: $\sin^22\theta_{12} =
0.0336$ and $\sin^22\theta_{13} = 0.0038$, which for $\delta
=0$ gives the same LSND result as $\sin^22\theta_{12} =
\sin^22\theta_{12} = 0.015$. The $CP$--violation effects for
$\theta_{12} \ne \theta_{13}$ are not as dramatic as with $\theta_{12} =
\theta_{13}$, but still may be observable with $10^{20}$~kt-decays.


An examination of Table~\ref{tab:CP} shows that the relative size of the
$CP$ violation in the $\nu_e \to \nu_\mu$ or $\nu_e \to \nu_\tau$
channels increases with $L/E$. However, once $\Delta _{\rm atm}$ is of
order unity or larger the $\sin2\Delta_{\rm atm}$ term in $S$
averages to zero, washing out the $CP$ violation; this does not happen
until $L$ is much larger than $100$~km. Since the flux falls off like
$1/L^2$, the statistical uncertainty increases roughly like
$L$, and as long as $\Delta_{\rm atm} \ll 1$
a wide range of distances have comparable sensitivity to
$CP$ violation in the $\nu_e \to \nu_\mu$ or $\nu_e \to \nu_\tau$
channels.

\begin{figure}
\centering\leavevmode
\epsfxsize=6.3in\epsffile{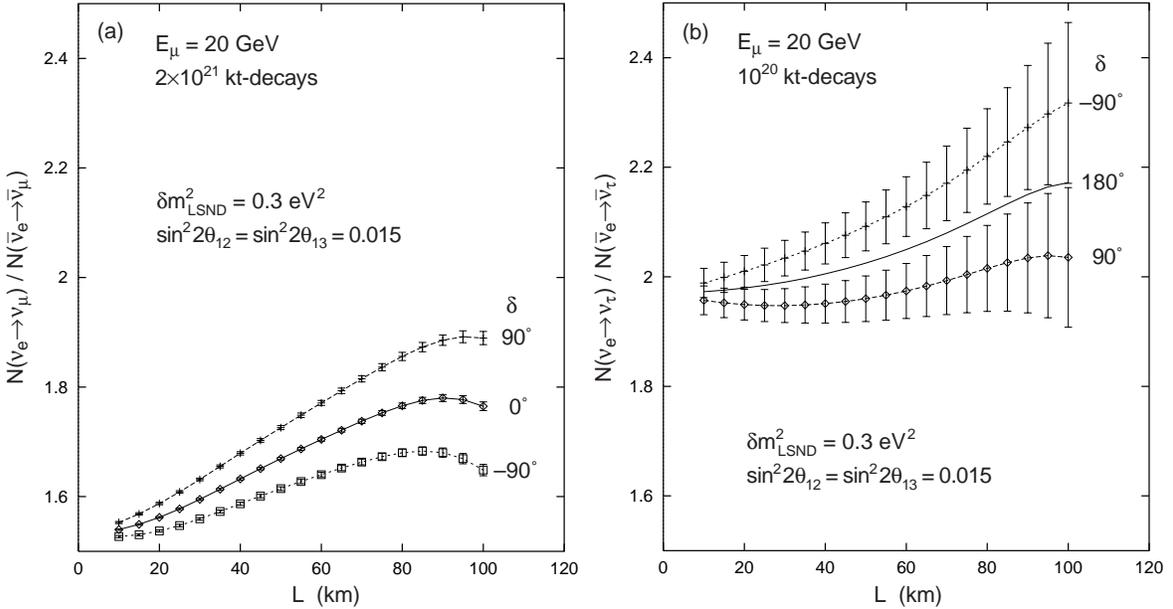}

\bigskip

\caption[]{(a) $R_{e\mu} \equiv N(\nu_e \to \nu_\mu)/N(\bar\nu_e \to
\bar\nu_\mu)$, and (b) $R_{e\tau} \equiv N(\nu_e \to
\nu_\tau)/N(\bar\nu_e \to \bar\nu_\tau)$.
A 4~GeV cut has been imposed on the detected muon in (a).
Statistical errors correspond to $2\times10^{21}$~kt-decays in (a)
and $10^{20}$~kt-decays in (b).}
\label{fig:e}
\end{figure}

Figure~\ref{fig:e}a shows the ratio $R_{e\mu}$ of $\nu_e \to \nu_\mu$
event rates (from $\mu^+$ decays) to $\bar\nu_e \to \bar\nu_\mu$ event
rates (from $\mu^-$ decays) versus baseline for oscillation parameters
given by Eq.~(\ref{eq:1B1}), with $E_\mu = 20$~GeV. The statistical
errors correspond to $2\times10^{21}$~kt-decays (which could be
obtained, for example, by three years
of running with $10^{20}$ useful muon decays per year, and a 10~kt
detector having a 67\% muon efficiency). A 4~GeV minimum energy cut has
been made on the detected muon. Although $R_{e\mu}$ is not as sensitive
to $CPV$ effects as $R_{\mu\tau}$, the increased statistics (due to a
larger overall rate resulting from the use of a larger detector for
muons) make $\nu_e \to \nu_\mu$ another attractive channel for $CP$
violation.  Figure~\ref{fig:e}b shows similar results for $R_{e\tau}$,
the ratio of $\nu_e \to \nu_\tau$ event rates (from $\mu^+$ decays) to
$\bar\nu_e \to \bar\nu_\tau$ event rates (from $\mu^-$ decays). The
statistical errors correspond to $10^{20}$~kt-decays.
It is evident from the figure that the $\nu_e \to \nu_\tau$ channel is
not as useful for detecting $CP$ violation (primarily because of the
reduced event rate in tau detection), although this channel is the
most sensitive to the $\nu_e \to \nu_\tau$ oscillation amplitude. The
combination of parameters $|s_{12}s_{23} - s_{13}c_{23}e^{i\delta}|$ in
the $\nu_e \to \nu_\tau$ amplitude is also present in the
$\sin^2\Delta_{\rm LSND}$ term of $P(\nu_\mu \to \nu_\tau)$ [see
Eq.~(\ref{eq:Pmt})], along with an additional factor involving small
mixing angles; hence, it would be more difficult to measure
$|s_{12}s_{23} - s_{13}c_{23}e^{i\delta}|$ in the $\nu_\mu \to \nu_\tau$
channel.

\begin{figure}
\centering\leavevmode
\epsfxsize=3.25in\epsffile{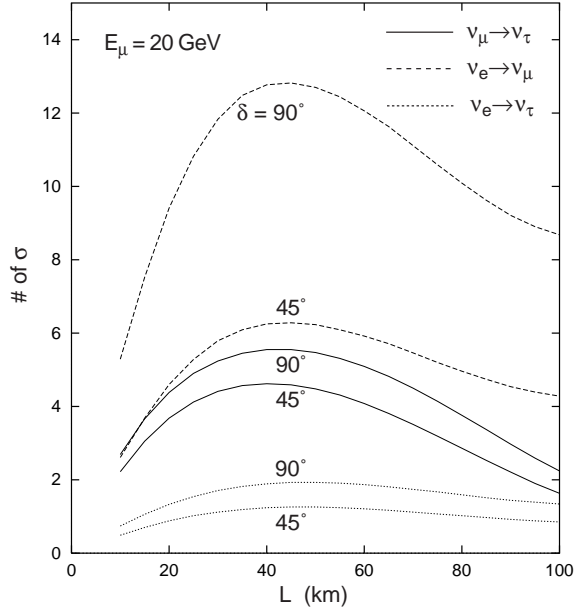}

\bigskip

\caption[]{Statistical significance of $CP$ violation measurements of
$R_{\mu\tau}$ (solid curves), $R_{e\mu}$ (dashed), and $R_{e\tau}$
(dotted) for various positive values of the $CP$-violating phase
$\delta$, when compared to the $CP$-conserving case, assuming $E_\mu =
20$~GeV, and $10^{20}$~kt-decays for a tau detection and
$2\times10^{21}$~kt-decays for muon detection. The
other oscillation parameters are given in Eq.~(\ref{eq:1B1}). Similar
results are obtained for negative values of $\delta$.}
\label{fig:signif}
\end{figure}

\begin{figure}
\centering\leavevmode
\epsfxsize=6.3in\epsffile{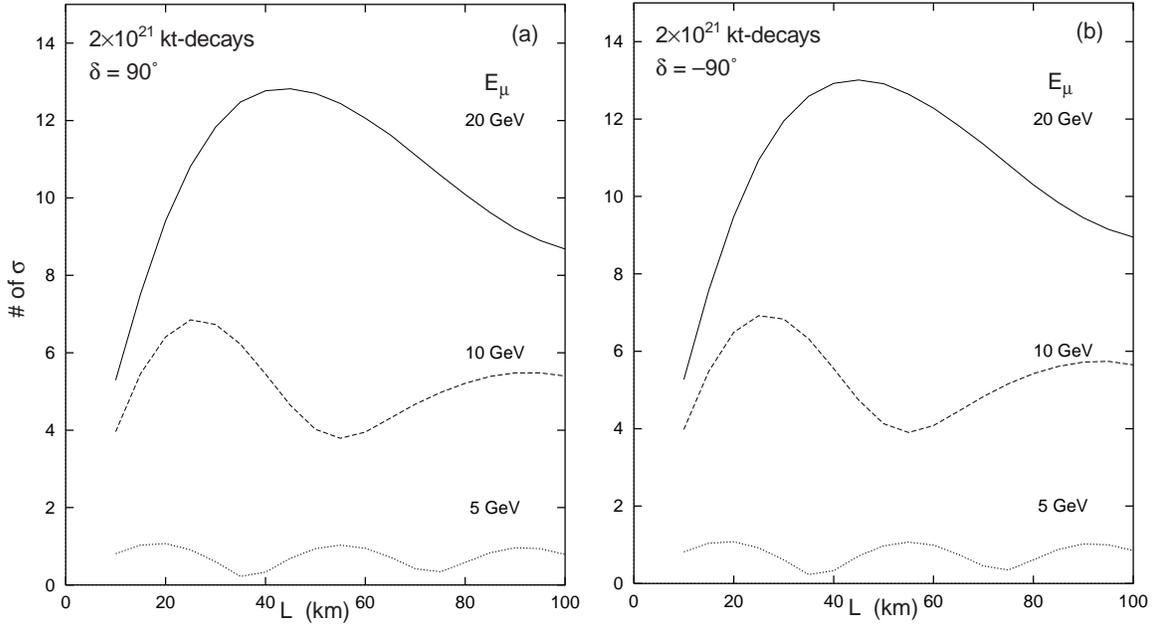}

\bigskip

\caption[]{Statistical significance of $CP$ violation in $R_{e\mu}$
for (a) $\delta = 90^\circ$ and (b) $\delta = -90^\circ$, when
compared
to the $CP$-conserving case, assuming $2\times10^{21}$~kt-decays,
for $E_\mu =$ 20~GeV (solid curve), 10~GeV (dashed), and 5~GeV (dotted).
The other oscillation parameters are given in Eq.~(\ref{eq:1B1}). A
4~GeV cut has been imposed on the detected muon.}
\label{fig:signif2}
\end{figure}

Fig.~\ref{fig:signif} shows, for various $CP$-violating cases with
$E_\mu = 20$~GeV, the statistical significance (number of standard
deviations) that the ratios $R_{\mu\tau}$, $R_{e\mu}$, and $R_{e\tau}$
deviate from their expected values for the $CP$-conserving case. While
both the $\nu_e \to \nu_\mu$ and $\nu_\mu \to \nu_\tau$ channels provide
good sensitivity near the optimal $L$, $\nu_e \to \nu_\mu$ is more
sensitive for a wider range of $L$, especially for values of $\delta$
that do not give maximal $CP$ violation. An additional potential
advantage of the $\nu_e \rightarrow \nu_\mu$ channel is that, in
principle, lower energy neutrino factories can be used since there is no
need to be above the tau-lepton production threshold. However, as the
energy of muons in a neutrino factory decreases, the sensitivity to CP
violation also decreases (see Fig.~\ref{fig:signif2}); this is
especially true if a lower bound is imposed on the energy of the
detected muon.

In principle, measurements can be made with varying $L/E$, either by
using experiments at more than one baseline, or by using a measure of
the neutrino energy (for example, the total observed event energy) at a
fixed baseline. However, since there are only five independent
quantities in the leading- and subleading-order probabilities in
Eqs.~(\ref{eq:Pem})--(\ref{eq:Pmt}), such measurements still cannot
completely determine the three-neutrino parameter set; this can only be
done at short baselines by a measurement of the subsubleading terms in
the off-diagonal probabilities, or of the subleading terms in the
diagonal $\nu_\mu \to \nu_\mu$ probability [see Eq~(\ref{eq:Pmm})], both
of which would be very challenging experimentally. Similar conclusions
apply for additional measurements involving electron detection.

To summarize, when used in conjunction with long-baseline measurements
from K2K, MINOS, ICANOE, and OPERA, and with results from MiniBooNE, all
six of the three-neutrino oscillation parameters can in principle be
determined with short-baseline measurements at a neutrino factory. The
short-baseline measurements would determine two parameters not
determined by other accelerator-based experiments, and would provide a
consistency check by independently measuring three other parameters also
measured by other accelerator-based experiments. The $\nu_e \to
\nu_\tau$ channel is most sensitive to the quantity $|s_{12}s_{23} -
s_{13}c_{23}e^{i\delta}|$.  The $\nu_e\to\nu_\mu$ channel provides good
sensitivity to $CP$ violation over a wide range of $L$ when a large muon
detector is used, and the $\nu_\mu\to\nu_\tau$ channel is most
useful for detecting $CP$ violation near the optimal $L$ for the
parameter choice illustrated.

\section{Four-neutrino models}

\subsection{Oscillation formalism}

Four-neutrino models are required to completely describe the solar,
atmospheric, and LSND data in terms of oscillations, since a third
independent mass-squared difference is necessary. A fourth neutrino must
be sterile, i.e., have negligible interactions, since only three
neutrinos are measured in $Z \to \nu\bar\nu$ decays~\cite{Znunubar}.
Following Ref.~\cite{4nuCP}, we label the fourth mass eigenvalue
$m_0$. Given the pattern of masses $m_1$, $m_2$, and $m_3$ from the
three neutrino case in Sec.~II, there is a preferred choice for the scale
of $m_0$ that can fit all of the data, including constraints from
accelerator experiments, namely, $m_0$ must be nearly degenerate with
$m_1$, so that there are two pairs of nearly degenerate states separated
by a mass gap of about 1~eV~\cite{bilenky,4nu}; see
Fig.~\ref{fig:levels}b. Then $\delta m^2_{10}$
governs the oscillation of solar neutrinos, $\delta m^2_{32}$ governs
the oscillation of atmospheric neutrinos, and $\delta m^2_{21}$, $\delta
m^2_{31}$, $\delta m^2_{20}$, and $\delta m^2_{30}$ all contribute to
the LSND oscillations.

Six mixing angles and three (six) phases are needed to parameterize the
mixing of four Dirac (Majorana) neutrinos; only three of these phases
are
measurable in neutrino oscillations. Thus three additional mixing
angles, which we label $\theta_{01}$, $\theta_{02}$, and $\theta_{03}$,
and two additional phases are required in extending the three-neutrino
phenomenology to the four neutrino case. The simplest situation,
which occurs in most explicit four-neutrino models, is that large
mixing occurs only between the nearly degenerate pairs; then the
four-neutrino mixing matrix can be parametrized as~\cite{4nuCP}
\begin{equation}
U = \left( \begin{array}{cccc}
c_{01} & s_{01}^* & s_{02}^* & s_{03}^*
\\
&&&\\
-s_{01} & c_{01} & s_{12}^* & s_{13}^*
\\
&&&\\
-c_{01}(s_{23}^*s_{03}+c_{23}s_{02})
& -s_{01}^*(s_{23}^*s_{03}+c_{23}s_{02})
& c_{23}
& s_{23}^*
\\
+s_{01}(s_{23}^*s_{13}+c_{23}s_{12})
& -c_{01}(s_{23}^*s_{13}+c_{23}s_{12})
&&\\
&&&\\
c_{01}(s_{23}s_{02}-c_{23}s_{03})
& s_{01}^*(s_{23}s_{02}c_{23}s_{03})
& -s_{23}
& c_{23}
\\
-s_{01}(s_{23}s_{12}-c_{23}s_{13})
& +c_{01}(s_{23}s_{12}-c_{23}s_{13})
&&\\
&&&\\
\end{array} \right) \,,
\label{genU2}
\end{equation}
where $s_{jk}$ is here defined as $\sin\theta_{jk}e^{i\delta_{jk}}$, and
the $\delta_{jk}$ are the six possible phases for Majorana neutrinos.
We set $\delta_{12} = \delta_{23} = \delta_{02} = 0$ without loss of
generality, since only three phases are measurable in neutrino
oscillations.

In this four-neutrino scenario, the parameters $\delta m^2_{32}$,
$\delta m^2_{21}$, $\theta_{23}$, $\theta_{12}$ and $\theta_{13}$ have
the same roles as in the three-neutrino scenario in the previous
section; the phase $\delta_{13}$ can be identified with $\delta$ in the
three-neutrino case. The angle $\theta_{01}$ describes the mixing of the
fourth flavor of neutrino with the state it is nearly degenerate with,
$\nu_e$; together $\delta m^2_{10}$ and $\theta_{01}$ can take on the
approximate values appropriate to any of the solar neutrino oscillation
solutions~\cite{solarfits} (small angle MSW, large angle MSW, LOW, or
vacuum). The remaining mixing angles $\theta_{02}$ and $\theta_{03}$
describe mixing of the fourth neutrino with the two states $\nu_\mu$,
$\nu_\tau$ in the other nearly-degenerate pair. Although models with
pure oscillations to sterile neutrinos are disfavored for the
solar~\cite{solarsterile} and atmospheric~\cite{atmossterile} data,
models with mixed oscillations to sterile and active neutrinos~\cite{4nu,sterile,sterile2},
i.e., non-negligible $\theta_{02}$ and $\theta_{03}$, are presumably
also possible. Finally, $\delta_{01}$ and $\delta_{03}$
are extra phases that may be observable in neutrino oscillations with
four neutrinos.

In the limit that terms involving the solar mass-squared difference can
be ignored, the expressions for the oscillation probabilities in
Eqs.~(\ref{eq:Pem}) and (\ref{eq:Pet}) remain the same. However, the
$\nu_\mu \to \nu_\tau$ probability becomes
\begin{eqnarray}
P(\nu_\mu \to \nu_\tau) &\simeq&
4|(s_{12}c_{23}+s_{13}s_{23}e^{i\delta_{13}})
  (s_{12}s_{23}-s_{13}c_{23}e^{-i\delta_{13}})
\nonumber\\
&\phantom{\simeq}&
+(s_{02}c_{23}+s_{03}s_{23}e^{i\delta_{03}})
 (s_{02}s_{23}-s_{03}c_{23}e^{-i\delta_{03}})|^2
\sin^2\Delta_{\rm LSND} \nonumber\\
&\phantom{\simeq}&
+ \sin^22\theta_{23}\sin^2\Delta_{\rm atm}
+ (s_{12}s_{13}\sin\delta_{13} + s_{02}s_{03}\sin\delta_{03})
\sin2\theta_{23} S \,.
\label{eq:Pmt2}
\end{eqnarray}
For $P(\bar\nu_\mu \to \bar\nu_\tau)$, the sign of the $S$-term
is reversed.

\subsection{Discussion}

When combined with measurements from
K2K, MINOS, ICANOE, OPERA, and MiniBooNE, the three parameters
$\theta_{12}$, $\theta_{13}$ and $\delta_{13}$ can in principle be
determined by short-baseline measurements of the $\nu_e \to \nu_\tau$
amplitude and the $CPV$ term in the $\nu_e \to \nu_\mu$ channel. This is
similar to the three-neutrino case in Sec.~II. However, measurements of
$\nu_\mu \to \nu_\tau$ and $\bar\nu_\mu \to \bar\nu_\tau$ will only give
partial information on $\theta_{02}$, $\theta_{03}$, and $\delta_{03}$.
The $\nu_\mu \to \nu_\mu$ survival probability can also be measured,
which in the limit that the solar mass-squared difference can be ignored
is
\begin{equation}
P(\nu_\mu \to \nu_\mu) \simeq 1 - 4 A (1-A) \sin^2\Delta_{\rm LSND}
-\sin^22\theta_{23}\sin^2\Delta_{\rm atm} \,,
\label{eq:Pmm2}
\end{equation}
where
\begin{equation}
A \simeq 4\left[ s_{23}^2(|s_{03}|^2+|s_{13}|^2)
+c_{23}^2(s_{02}^2 + s_{12}^2) + 2 s_{23}c_{23} {\rm Re}(s_{02}s_{03}
+s_{12}s_{13}) \right] \,.
\end{equation}
Both oscillatory terms in Eq.~(\ref{eq:Pmm2}) are second order in small
quantities in short-baseline experiments. In principle $P(\nu_\mu \to
\nu_\mu)$ could be used as an additional measurement; however, unless
$\theta_{02}$ and $\theta_{03}$ are larger than $\theta_{12}$ and
$\theta_{13}$, the deviations of $P(\nu_\mu \to \nu_\mu)$ from unity are
very small, and the normalization of the beam flux would have to be
known to high precision for this to be useful.


The parameters $s_{12}$, $s_{13}$ and $s_\delta$ determined by
short baseline measurements of the $\nu_e \to \nu_\tau$ and $\nu_e \to
\nu_\mu$ channels give predictions for the $\nu_\mu \to \nu_\tau$
channel at short baselines that can differ for the three- and
four-neutrino cases [Eqs.~(\ref{eq:Pmt}) and (\ref{eq:Pmt2}),
respectively]. If the three-neutrino predictions for $\nu_\mu \to
\nu_\tau$ are found to substantially disagree with the experimental
measurements, then the disagreement would provide evidence for the
existence of a fourth neutrino. The absence of
such a disagreement would indicate that either there are only three
neutrinos or that $\theta_{02}$ and $\theta_{03}$ are significantly
smaller than $\theta_{12}$ and $\theta_{13}$.

\begin{figure}
\centering\leavevmode
\epsfxsize=6.3in\epsffile{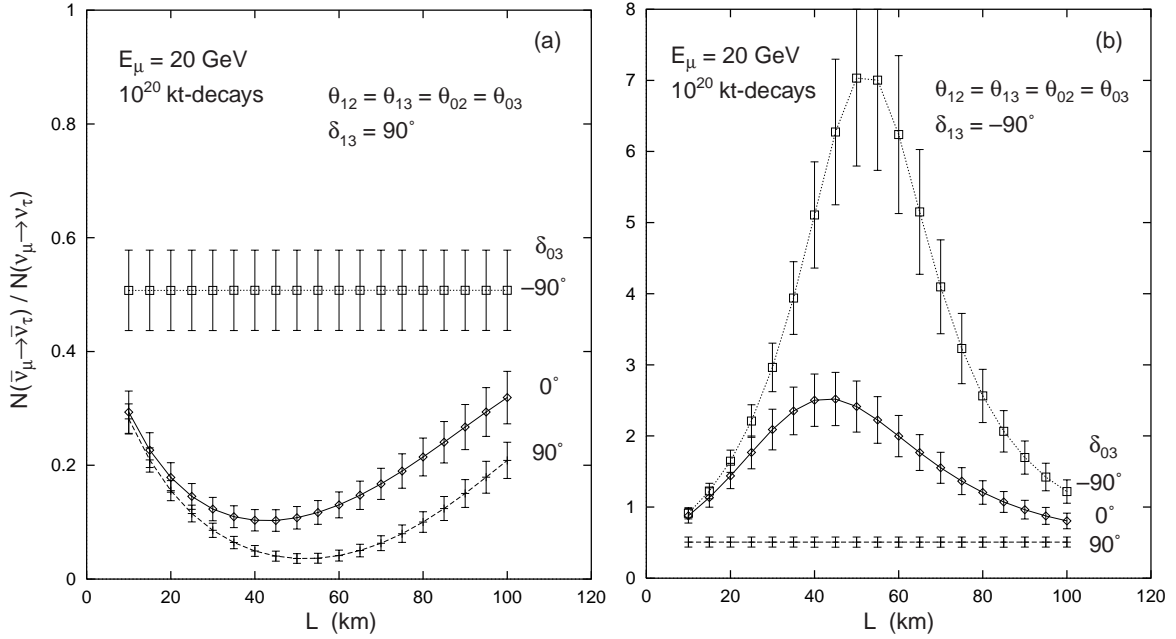}

\bigskip

\caption[]{$CP$-violation effects with four neutrinos in
$R_{\mu\tau}$ versus $L$, for $E_\mu = 20$~GeV with oscillation
parameters given by Eq.~(\ref{eq:1B1}), $\theta_{02} = \theta_{03} =
\theta_{12} = \theta_{13}$, $\delta_{03} = 0$ (solid curves), $90^\circ$
(dashed), and $-90^\circ$ (dotted) and (a) $\delta_{13} = 90^\circ$ and
(b) $\delta_{13} = -90^\circ$. Statistical errors correspond to
$10^{20}$~kt-decays. Corresponding results for three neutrinos are given
in Fig.~\ref{fig:mutau}.}
\label{fig:4nu}
\end{figure}

\begin{figure}
\centering\leavevmode
\epsfxsize=3.25in\epsffile{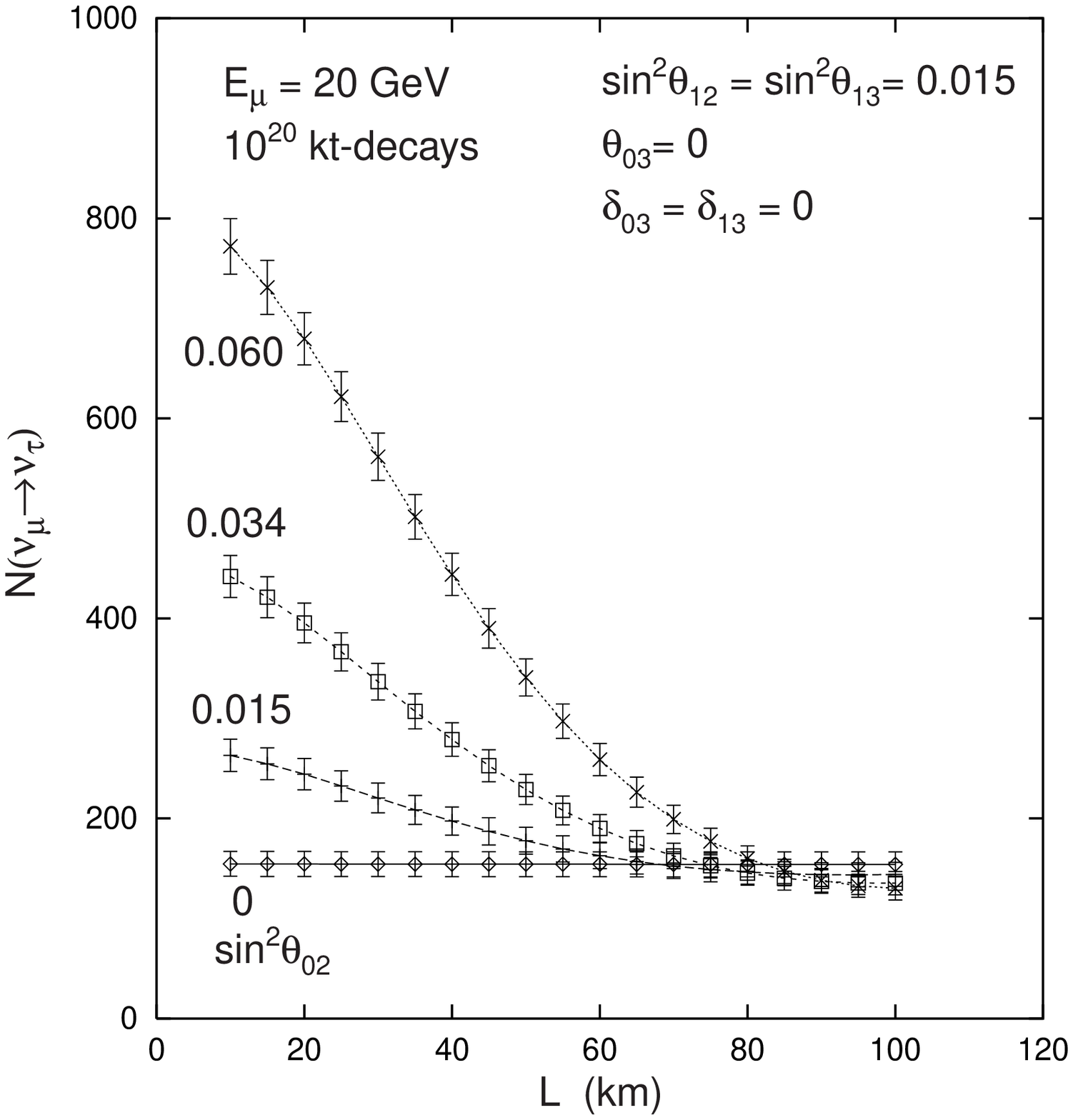}

\bigskip

\caption[]{Effects of four-neutrino mixing on the $\nu_\mu \to
\nu_\tau$ event rate versus $L$
for $E_\mu = 20$~GeV, $\theta_{03} =
\delta_{13} = \delta_{03} = 0$, $\theta_{02} = \theta_{03}$, and
$\sin^22\theta_{02} = 0$ (solid curve), $0.015$ (dashed), $0.034$
(dotted), and $0.060$ (dot-dashed). Statistical errors correspond to
$10^{20}$~kt-decays.}
\label{fig:4nu2}
\end{figure}

In general, the sensitivity of the measurements of $\nu_\mu \to
\nu_\tau$ and $\bar\nu_\mu \to \bar\nu_\tau$ to $\theta_{02}$,
$\theta_{03}$, and $\delta_{03}$ in the four-neutrino case are similar
to the sensitivities to $\theta_{12}$, $\theta_{13}$, and $\delta$ in
the three-neutrino case. For example, given the parameters in
Eq.~(\ref{eq:1B1}) and $\delta_{13}=0$ (i.e., no $CP$
violation in the $\nu_e \to \nu_\mu$ and $\nu_e \to \nu_\tau$
channels), if $s_{02} = s_{12}$ and
$s_{03} = s_{13}$, then the four-neutrino predictions would be given by
Figs.~\ref{fig:mutau}a and \ref{fig:mutau}b, where $\delta_{03}$ takes
on the values of $\delta$ in the figures; the corresponding
three-neutrino predictions would be given by the $\delta=0$
curves. Larger values of $s_{02}$ and $s_{03}$ could give a much larger
$CP$-violating effect, provided that $\delta_{03}$ was not small. If
both $\delta_{03}$ and $\delta_{02}$ were nonzero, their $CP$-violating
effects could add together either constructively or destructively.
Hence, larger $CP$-violating effects are possible
in the $\nu_\mu \to \nu_\tau$ channel in the four-neutrino case than
with three neutrinos, or
$CP$ violation may be present in $\nu_\mu \to \nu_\tau$ when it is not
present in the $\nu_e \to \nu_\mu$ or $\nu_e \to \nu_\tau$ channels (or
vice versa), unlike the three-neutrino case, as illustrated in
Fig.~\ref{fig:4nu}. Even if there is no $CP$
violation, effects of the angles $\theta_{02}$ and $\theta_{03}$ could
be seen in the $\sin^2\Delta_{\rm LSND}$ term in Eq.~(\ref{eq:Pmt2}),
as illustrated in Fig.~\ref{fig:4nu2}. Here, sensitivities to $\theta_{02}$
and $\theta_{03}$ tend to increase with decreasing $L$, due to the
increased flux at shorter distances.

Proof of the existence of a fourth neutrino does not exclude the
possibility that there may be more than four neutrinos. Strictly
speaking, an inconsistency between the measurement of $\nu_\mu \to
\nu_\tau$ oscillations and the three-neutrino predictions would imply
only that there are four or more neutrinos. In a model with four or more
neutrinos, there are many more mixing angles and phases in the neutrino
mixing matrix, and it would not be possible to determine them all from
these oscillation measurements. However, when combined with results at
other baselines, most of the four-neutrino parameter set could be
determined~\cite{4nuCP}.

\section{Summary}

In scenarios designed to describe the LSND oscillation results, our
results show that short-baseline neutrino factory measurements can in
principle determine all of the three-neutrino parameters provided the
results are used together with future results from other
accelerator-based experiments.  The $\nu_e \to \nu_\tau$ channel is most
sensitive to one combination of parameters in the $CP$-conserving
terms. The $\nu_e \to \nu_\mu$ channel provides good sensitivity to $CP$
violation over a wide range of $L$ (20--100~km for $E_\mu = 20$~GeV),
assuming that a large muon detector is used, e.g.\ 10~kt. The $\nu_\mu
\to \nu_\tau$ channel is also sensitive to $CP$ violation for a more
restricted range of $L$, and may be used to explore whether more than
three neutrinos exist. For four or more neutrinos, the $CP$-violating
effect in $\nu_\mu \to \nu_\tau$ may be either enhanced or reduced by
the additional mixing parameters.
If MiniBooNE confirms the LSND oscillation results, then it will
be important to measure the rates in all three appearance modes, as
well as to search for $CP$ violation in $\nu_e \rightarrow \nu_\mu$ and
$\nu_\mu \rightarrow \nu_\tau$, to obtain indirect evidence for the
existence of sterile neutrinos. Neutral current measurements would
complement the charge current studies of this paper in the search for
sterile neutrinos.

\acknowledgments
VB thanks the Aspen Center for Physics for hospitality during the course of this work.
This research was supported in part by the U.S.~Department of Energy
under Grant Nos.~DE-FG02-94ER40817, DE-FG02-95ER40896 and
DE-AC02-76CH03000, and in part by the University of Wisconsin Research
Committee with funds granted by the Wisconsin Alumni Research
Foundation.

\newpage

\begin{thebibliography}{99}

\bibitem{SuperKsolar}
Super-Kamiokande collaboration, Y. Fukuda {\it et al.},
Phys. Rev. Lett. {\bf 81}, 1158 (1998);
{\bf 82}, 1810 (1999);
{\bf 82}, 2430 (1999).

\bibitem{SuperKatmos}
Super-Kamiokande Collaboration, Y. Fukuda {\it et al.},
Phys. Lett. {\bf B433}, 9 (1998); {\bf B436}, 33 (1998);
Phys. Rev. Lett. {\bf 81}, 1562 (1998); {\bf 82}, 2644 (1999).
Phys. Lett. {\bf B467}, 185 (1999);

\bibitem{LSND}
C. Athanassopoulos et al. (LSND Collab.),
Phys. Rev. Lett. {\bf 77}, 3082 (1996); {\bf 81}, 1774 (1998);
G. Mills, talk at {\it Neutrino-2000}, XIXth International Conference
on Neutrino Physics and Astrophysics, Sudbury, Canada, June 2000.

\bibitem{othersolar}
B.T. Cleveland {\it et al.},
Nucl. Phys. B (Proc. Suppl.) {\bf 38}, 47 (1995);
GALLEX collaboration, W. Hampel {\it et al.},
Phys. Lett. {\bf B388}, 384 (1996);
SAGE collaboration, J.N. Abdurashitov {\it et al.},
Phys. Rev. Lett. {\bf 77}, 4708 (1996);
Kamiokande collaboration, Y. Fukuda {\it et al.},
Phys. Rev. Lett, {\bf 77}, 1683 (1996).

\bibitem{other}
S.~Pakvasa, hep-ph/9905426, in {\it Proc. of the 8th International
Conference on Neutrino Telescopes}, Venice, Feb.~1999, ed. by
M.~Baldo-Ceolin (Papergraf, Padova, 1999), vol.~1, p.~283;
E.~Roulet, Phys. Rev. {\bf D44}, 935 (1991);
M.M.~Guzzo, A.~Masiero and S.~Petcov, Phys. Lett. {\bf B260}, 154 (1991);
V.~Barger, R.J.N.~Phillips, and K.~Whisnant,
Phys. Rev. {\bf D44}, 1629 (1991);
S.~Coleman  and S.L.~Glashow, Phys. Lett. {\bf B405}, 249 (1997);
J.N.~Bahcall and P.~Krastev, hep-ph/9703267, in {\it Proc. of the
Symposium on Flavor Changing Neutral Currents (FCNC\,97)}, Santa Monica,
Feb.~1997, ed. by D.~Cline (World Scientific, Singapore, 1997), p.~259;
S.~Bergmann, M.M.~Guzzo, P.C.~de Holanda, P.~Krastev, and H.~Nunokawa,
hep-ph/0004049.

\bibitem{bilenky}
S.M. Bilenky, C. Giunti, and W. Grimus,
Eur. Phys. J. {\bf C1}, 247 (1998);

\bibitem{4nu}
V. Barger, K. Whisnant, and T.J. Weiler,
Phys. Lett. {\bf B427}, 97 (1998);
V. Barger, S. Pakvasa, T.J. Weiler, and K. Whisnant,
Phys. Rev. {\bf D58}, 093016 (1998).

\bibitem{nufact}
S. Geer, Phys. Rev. {\bf D57}, 6989 (1998);
see also the Fermilab Long-Baseline Workshop web site at
http://www.fnal.gov/projects/muon\_collider/nu/study/study.html

\bibitem{derujula}
A. De Rujula, M. B. Gavela, and P. Hernandez,
Nucl. Phys. {\bf B547}, 21 (1999).

\bibitem{nufact2}
B. Autin et al., ``Physics opportunities at a
CERN based neutrino factory", CERN-SPSC-98-30, Oct. 1998; 
A. Bueno, M. Campanelli, A. Rubbia, hep-ph/9808485; hep-ph/9809252;
B. Autin, A. Blondel, J. Ellis
(editors), ``Prospective Study of Muon Storage Rings at CERN", CERN
99-02, April 1999; 
and A. Blondel {\it et al.}, ``The Neutrino Factory: Beam and
Experiments,'' CERN-EP-2000-053, April 2000.

\bibitem{BGW}
V. Barger, S. Geer, and K. Whisnant,
Phys.\ Rev. {\bf D 61}, 053004 (2000).

\bibitem{BGRW}
V. Barger, S. Geer, R. Raja, and K. Whisnant,
Phys. Rev. {\bf D 62}, 013004 (2000); hep-ph/0003184, to be published in Phys. Rev.~D.

\bibitem{cervera}
A. Cervera, A. Donini, M.B. Gavela, J.J. Gomez Cadenas, P. Hernandez,
O. Mena, and S. Rigolin, Nucl.Phys. {\bf B 579}, 17 (2000).

\bibitem{rubbia}
M. Campanelli, A. Bueno, and A. Rubbia, hep-ph/0005007.

\bibitem{freund}
I. Mocioiu and R. Shrock, hep-ph/0002149;
M. Freund, P. Huber, and M. Lindner, hep-ph/0004085.

\bibitem{othernufact}
M. Freund, M. Lindner, S.T. Petcov, and M. Romanino,
Nucl. Phys. {\bf B 578}, 29 (2000);
M. Campanelli, A. Bueno, and A. Rubbia,
Nucl. Phys. {\bf B 573}, 27 (2000).

\bibitem{mns}
Z. Maki, M. Nakagawa, and S. Sakata,
Prog. Theor. Phys. {\bf 28}, 870 (1962).

\bibitem{keung}
W.-Y. Keung and L.-L. Chau, Phys. Rev. Lett. {\bf 53}, 1802 (1984).

\bibitem{jarlskog}
C. Jarlskog, Z. Phys. {\bf C 29}, 491 (1985); Phys. Rev. {\bf D 35},
1685 (1987).

\bibitem{4nuCP}
V. Barger, Y.-B. Dai, K. Whisnant, and B.-L. Young,
Phys. Rev. {\bf D59}, 113010 (1999).

\bibitem{donini}
A. Donini, M.B. Gavela, P. Hernandez, and S. Rigolin,
Nucl. Phys. {\bf B 574}, 23 (2000).

\bibitem{CPV}
D.J. Wagner and T.J. Weiler, Phys. Rev. {\bf D59}, 113007 (1999);
A.M.~Gago, V.~Pleitez, R.Z.~Funchal, hep-ph/9810505;
K.R.~Schubert, hep-ph/9902215;
K.~Dick, M.~Freund, M.~Lindner and A.~Romanino, hep-ph/9903308;
A.~Romanino, hep-ph/9909425;
J.~Bernabeu, hep-ph/9904474;
S.M.~Bilenky, C.~Giunti, and W.~Grimus,
Phys. Rev. {\bf D58}, 033001 (1998);
M.~Tanimoto, Prog. Theor. Phys. {\bf 97}, 901 (1997);
J.~Arafune, J.~Sato, Phys. Rev. {\bf D55}, 1653 (1997);
T.~Fukuyama, K.~Matasuda, H.~Nishiura,
Phys. Rev. {\bf D57}, 5844 (1998);
M.~Koike and J.~Sato, hep-ph/9707203,
Proc. of the KEK Meetings on CP Violation and its Origin;
H.~Minakata and H.~Nunokawa, Phys. Lett. {\bf B413}, 369 (1997);
H.~Minakata and H.~Nunokawa, Phys. Rev. {\bf D57}, 4403 (1998);
J.~Arafune, M.~Koike and J.~Sato, Phys. Rev. {\bf D56}, 3093 (1997);
M.~Tanimoto, Phys. Lett. {\bf B435}, 373 (1998).

\bibitem{matter}
L. Wolfenstein, Phys. Rev. {\bf D17}, 2369 (1978);
V.~Barger, S.~Pakvasa, R.J.N.~Phillips, and K.~Whisnant, Phys.
Rev. {\bf D22}, 2718 (1980);
P. Langacker, J.P. Leveille, and J. Sheiman,
Phys. Rev. {\bf D 27}, 1228 (1983);
S.P. Mikheyev and A. Smirnov, Yad. Fiz. {\bf 42}, 1441 (1985)
[Sov. J. Nucl. Phys. 42, 913 (1986)].

\bibitem{BUGEY}
Y. Declais {\it et al.}, Nucl. Phys. {\bf B434}, 503 (1995).

\bibitem{1B1}
C.~Albright et al., ``Physics at a Neutrino Factory",  FERMILAB-FN-692,
Sec.~3.1, http://www.fnal.gov/projects/muon\_collider/nu/study/report.ps

\bibitem{K2K}
K. Nishikawa et al. (KEK-PS E362 Collab.),
``Proposal for a Long Baseline Neutrino Oscillation Experiment,
using KEK-PS and Super-Kamiokande", 1995, unpublished.

\bibitem{MINOS}
MINOS Collaboration, ``Neutrino Oscillation Physics at Fermilab: The
NuMI-MINOS Project,'' NuMI-L-375, May 1998.

\bibitem{ICANOE}
See the ICANOE web page at http://pcnometh4.cern.ch/

\bibitem{OPERA}
See the OPERA web page at http://www.cern.ch/opera/

\bibitem{MiniBooNE}
E. Church et al. (BooNE Collab.), ``A letter of intent
for an experiment to measure $\nu_\mu \rightarrow \nu_e$
oscillations and $\nu_\mu$  at the Fermilab Booster",
May 16, 1997, unpublished.

\bibitem{Znunubar}
LEP Electroweak Working Group and SLD Heavy Flavor Group,
D. Abbaneo {\it et al.}, CERN-PPE-96-183, December 1996.

\bibitem{solarfits}
J.N. Bahcall, P.I. Krastev, and A.Yu. Smirnov,
Phys, Rev. {\bf D 58}, 096016 (1998);
M.C. Gonzalez-Garcia, P.C. de Holanda, C. Pena-Garay, and
J.W.F. Valle, Nucl. Phys. {\bf B 573}, 3 (2000);
G.L. Fogli, E. Lisi, and D. Montanino,
Astropart. Phys. {\bf 9}, 119 (1998).

\bibitem{solarsterile}
Y. Suzuki, Super-Kamiokande Collaboration, talk at {\it Neutrino-2000},
XIXth International Conference on Neutrino Physics and Astrophysics,
Sudbury, Canada, June 2000;
see also the first two papers of Ref.~\cite{solarfits}.

\bibitem{atmossterile}
H. Sobel, Super-Kamiokande Collaboration, talk at {\it Neutrino-2000},
XIXth International Conference on Neutrino Physics and Astrophysics,
Sudbury, Canada, June 2000;
see also N. Fornengo, M.C. Gonzalez-Garcia, and J.W.F. Valle,
Nucl. Phys. {\bf 580}, 58 (2000).

\bibitem{sterile}
Q.Y. Liu and A. Yu. Smirnov, Nucl. Phys. {\bf B 524}, 505 (1998).

\bibitem{sterile2}
C. Giunti, M.C. Gonzalez-Garcia, and C. Pena-Garay, hep-ph/0001101;
O. Yasuda, hep-ph/0006319;
E. Lisi, talk at {\it Neutrino-2000}, XIXth International Conference on
Neutrino Physics and Astrophysics, Sudbury, Canada, June 2000.


\end{thebibliography}
\end{document}